\documentclass[onecolumn,showpacs,superscriptaddress]{revtex4}

\usepackage{amsfonts,xfrac}
\usepackage{amsmath}
\usepackage{amssymb}
\usepackage{graphicx}
\usepackage{dcolumn}
\usepackage{dsfont}
\usepackage{times}
\usepackage{epsfig,color}
\usepackage[]{units}

\newcommand{\R}{{\mathbb R}}

\newcommand{\Z}{{\mathbb Z}}

\newcommand{\C}{{\mathbb C}}

\newcommand{\pa}{\partial}

\def\epsilon{\varepsilon}
    
\def\beq{\begin{equation}}
\def\eq{\end{equation}}

\newcommand{\subfigimg}[3][,]{%
  \setbox1=\hbox{\includegraphics[#1]{#3}}
  \leavevmode\rlap{\usebox1}
  \rlap{\hspace*{5pt}\raisebox{\dimexpr\ht1-.5\baselineskip}{#2}}
  \phantom{\usebox1}
}

\begin{document}

\title{Time- and space-variant wave transmission in helicoidal phononic crystals}

\author{F. Li}
\affiliation{Aeronautics and Astronautics, University of Washington,
Seattle, WA 98195-2400}
\author{C. Chong$^*$}
\affiliation{Department of Mathematics and Statistics, University of Massachusetts,
Amherst MA 01003-4515, USA}
\affiliation{Department of Mechanical and Process Engineering (D-MAVT), 
Swiss Federal Institute of Technology (ETH),
8092
Zurich, Switzerland}
\author{J. Yang$^*$}
\affiliation{Aeronautics and Astronautics, University of Washington,
Seattle, WA 98195-2400}
\author{P. G. Kevrekidis}
\affiliation{Department of Mathematics and Statistics, University of Massachusetts,
Amherst MA 01003-4515, USA}
\author{C. Daraio}
\affiliation{Department of Mechanical and Process Engineering (D-MAVT), 
Swiss Federal Institute of Technology (ETH),
8092
Zurich, Switzerland}

\date{\today}

\begin{abstract}
 We present a dynamically tunable mechanism of wave transmission in 1D helicoidal
phononic crystals in a shape similar to DNA structures. These helicoidal architectures
allow slanted nonlinear contact among cylindrical constituents, 
and the relative torsional movements can dynamically tune the contact 
stiffness between neighboring cylinders. 
This results in cross-talking
between in-plane torsional and out-of-plane longitudinal waves. We numerically
demonstrate their versatile wave mixing and controllable 
dispersion behavior in both wavenumber and
frequency domains. Based on this principle, a
suggestion towards an acoustic configuration bearing 
parallels to a transistor is further proposed, in
which longitudinal waves can be switched on/off through torsional waves.
\end{abstract}

\maketitle

\section{Introduction}
Phononic crystals (PCs) are spatially periodic structures which can manipulate
acoustic waves more effectively compared to natural materials \cite{Sanchez,Liu}.  Intense recent research efforts along this direction have shown
that the acoustic characteristics of PCs depend on their 
given material properties, geometrical
configurations, and boundary conditions. For example, the frequency band structures of PCs can
be modified by changing their structural compositions or by the application of external fields \cite{Khelif,Lin,Jun}. Although tunable PCs have been investigated both theoretically and experimentally, their
acoustic properties are typically fixed by their initial design parameters and are not allowed to
vary adaptively, which limits the breadth of their potential applications. Previous studies explored the possibility of
altering wave transmission characteristics \emph{in-situ} by using time-varying material properties \cite{Wright}
and by exploiting amplitude-dependent responses of their nonlinear constituents \cite{Narisetti,Yang,bertoldi}. However,
dynamically tunable PCs are relatively unexplored, and wave propagation mechanisms in
time- and space-variant PCs remain largely unknown.

Recently the specific paradigm of granular crystals 
based on the assembly of discrete particles has attracted
significant attention due to their nonlinearity stemming from Hertzian contact \cite{Nester01,Sen}. The tunability of this nonlinearity from the weakly
to the highly nonlinear regime involves a degree of freedom that is
significant in this regard~\cite{Nester01,Sen,pgk_review,Theocharis_rev,Herbold09}.
Particularly, PCs with cylindrical elements have shown their dynamic versatility in controlling
the speed of nonlinear waves \cite{Khatri} and manipulating the cutoff frequencies of band gaps over
remarkably wide ranges \cite{Li}. These studies leveraged the variations of contact stiffness among
slanted cylinders by changing their alignment angles in a static manner.

In the present work, we report on the dynamic manipulation of wave propagation modes in one-dimensional
(1D) PCs made of helically stacked cylinders defined as helicoidal phononic
crystals (HPCs) herein. We impose in-plane torsional waves to 
HPCs by systematically
perturbing the alignment angles of cylinders in the 
temporal and spatial domains. 
This results in
dynamic variations of axial contact stiffness in the helicoidal structures, thereby making
longitudinal waves coupled with the torsional waves. In principle, this phonon-phonon scattering
(i.e., wave mixing) effect is equivalent to optical Brillouin 
scattering~\cite{brillouin}. Such dynamic cross-talking
between torsional and longitudinal waves 
offers an unprecedented controllability over wave
transmission in PCs, exhibiting fundamentally distinct
characteristics in comparison to conventional PCs with a fixed 
landscape of wave
dispersions. Herein, we demonstrate, for the first time, versatile manipulation
of wave dispersion mechanisms of a 
certain wave mode via another by using time- and space-variant
HPCs. In particular, we employ three distinct schemes: one in which the
HPCs are space-independent, varying solely with time; one in which
they are time-independent but are varying with space, and one in which both
the tunability in space and that in time are employed concurrently.
Finally, although acoustic diodes have been proposed~\cite{diode1,diode2}, 
we note that an acoustic transistor has not been realized. Based on HPCs, 
we propose a configuration bearing characteristics of a transistor, 
in that a longitudinal wave can be controlled actively by a torsional wave.

Our presentation is structured as follows: In Sec.~\ref{sec:model}, we analyze
the general model and give details of the availability of parametric
variations. In Sec.~\ref{sec:time} we consider time-dependent, space-independent
helicoidal configurations, in Sec.~\ref{sec:space} time-independent but
spatially dependent ones are studied and in Sec.~\ref{sec:spacetime} we combine both variations.
In Sec.~\ref{sec:transistor}, we provide a potential application of the HPCs with
a view towards a future implementation of an acoustic transistor.
Finally, in Sec.~\ref{sec:theend}, we summarize our findings and provide 
some directions for future study.


\section{Model}\label{sec:model}

The HPC investigated in this study is arranged in a DNA-like helical architecture with
increasing alignment angles 
(we denote the absolute angle of the $n$-th particle with respect to the first particle as $\alpha_n$).
The schematic of the HPC is shown in the inset of 
Fig.~\ref{fig:stiffness}(a). 
Each cylindrical element supports both
longitudinal and rotational movements. Longitudinal waves propagate along the helical chain
through the point contact among neighboring elements, 
 while torsional waves are imposed on the HPC at will by actuating each one of the cylinders independently.
 We neglect the torsional elasticity or friction at the contact.
 The relative angle between neighboring cylinder particles is  $\Delta \alpha_n = | \alpha_n - \alpha_{n-1}|$.
The contact force between two adjacent cylindrical particles is expressed by the Hertzian
law $F = k_{cyl}(\Delta \alpha) \delta^{3/2}$ for $\alpha \neq 0$, where $F$ and $\delta$ are the contact force and displacement, respectively. The
contact stiffness coefficient is given as a function of the angle
of other relevant material parameters as~\cite{Johnson}:
$$ k_{cyl}(\Delta \alpha) = \frac{\pi \sqrt{2} Y }{ 3( 1 - \nu^2) e K(e)^{3/2}} \left(    \left[    \left(\frac{1}{1-e^2}\right)E(e) - K(e)       \right][ K(e) - E(e)  ]           \right)^{1/4}. $$
 Here $K(e)$ and $E(e)$ are the complete elliptic integrals of the first and second kinds, respectively, and
 $e = \sqrt{1 - [\cos(\Delta \alpha ) / (1+\cos(\Delta\alpha )) ]^{4/3 } }$ is the eccentricity of the elliptical contact area between two
 cylindrical particles. Furthermore, $Y$ represents the Young's
modulus and $\nu$ the Poisson ratio. The contact stiffness $k_{cyl}$ is sensitive to the relative alignment angle $\Delta \alpha$ between
 adjacent cylindrical particles, implying that the dynamics of the axial and rotational motions of
 HPCs are coupled.  The longitudinal motion of the HPC can be written in terms of both rotational angle $\Delta\alpha_n$ and
axial displacement $u_n$ of the $n$-th element:
\begin{equation}\label{eq:eomNL}
 M \, \ddot{u}_n = k_{cyl}(\Delta \alpha_{n}) [ \delta_{n} + u_{n-1} - u_n ]_+^{3/2} -  k_{cyl}(\Delta \alpha_{n+1}) [ \delta_{n+1} + u_{n} - u_{n+1} ]_+^{3/2}   
\end{equation}
where $M$ is the mass of a cylindrical unit and $\delta_n$ is the deformation at static equilibrium between
the $n$-th and $n+1$-st particles resulting from the precompression force $F_0 = k_{cyl}(\Delta \alpha_{n}) \delta_{n}^{3/2}$. The bracket is defined by $[x]_+ = \mathrm{max}(0, x)$, denoting that there are
no tensional forces among cylinders. 
For the numerical results reported in this study, we consider HPCs composed of 200 fused quartz cylinders with diameter 18.0 mm and height 18.0 mm. The
mass, Young's modulus, and Poisson's ratio are
$M = 10$ g, $Y = 72$ GPa, and $\nu = 0.17$,
respectively.
The precompression ($F_0$) of the chain is assumed to be 20 N. 
The dimensions, material properties, and boundary conditions are based on the parameters used in our previous experimental studies \cite{Khatri}.

For small relative displacements $|u_{n-1}-u_{n}| \ll \delta_n$, the stiffness of the Hertzian contact can be
linearized \cite{Nester01}. Thus, we can infer dispersion properties of this system by studying the linearized equations of motion
\begin{equation}\label{eq:eom}
M \ddot{u}_n = k_{lin}(\Delta \alpha_{n}) u_{n-1} - (k_{lin}(\Delta \alpha_{n}) + k_{lin}(\Delta \alpha_{n+1}))u_{n} + k_{lin}(\Delta \alpha_{n+1})  u_{n+1}
\end{equation}
where the linear stiffness coefficient is given as
$k_{lin}(\Delta \alpha) = 3/2 \, k_{cyl}(\Delta \alpha)  \delta_0^{1/2}$.
While the nonlinear case of Eq.~(\ref{eq:eomNL}) is extremely
interesting in its own right and we will return to it in Sec.~\ref{sec:transistor},
for the purposes of the present work, we will restrict most of our considerations
to the linearized problem, i.e. Eq.~(\ref{eq:eom}).

We first consider as a reference case a regular HPC without 
introducing any dynamic perturbations of
angles (i.e., the standard, homogeneous case). 
More specifically, cylindrical angles in the helicoidal chain 
increase linearly by $\alpha_0$ (i.e.,
$\alpha_n = n \alpha_0$), and thus the relative angles between neighboring particles remain constant ($\Delta\alpha = \alpha_0$). Then,
the axial stiffness $k_{cyl}$ (and hence $k_{lin}$) is the same along the chain, which is equivalent to a standard, homogeneous
chain \cite{Nester01}. In this instance, the linear problem~\eqref{eq:eom} is solved by means of the Fourier mode 
plane wave ansatz $u_n = \exp(ikn + i \omega t)$
where the wavenumber $k$ and the frequency $\omega$ satisfy the dispersion 
relation
\begin{equation} \label{eq:disp_hom}
  \omega(k)^2 = 2 k_{lin}(\alpha_0)(1- \cos(k))/M 
\end{equation}
such that the maximum allowable frequency (i.e. the cutoff frequency) is $f_{\rm cutoff} = \sqrt{\frac{k_{lin}}{M}}/\pi$.
The computation of dispersion relations in the case of dynamic HPCs is 
considerably more involved. Thus, we now turn, in 
Secs.~\ref{sec:time}-\ref{sec:spacetime}, to the reformulation of
results on the spectra of spatially and/or temporally periodic
linear difference operators and compare these semi-analytical results to full 
numerical simulations of the nonlinear
model~\eqref{eq:eomNL} in the case of small relative displacements. 

%


\section{Time-variant, space-independent HPCs} \label{sec:time}

 \begin{figure}[t]
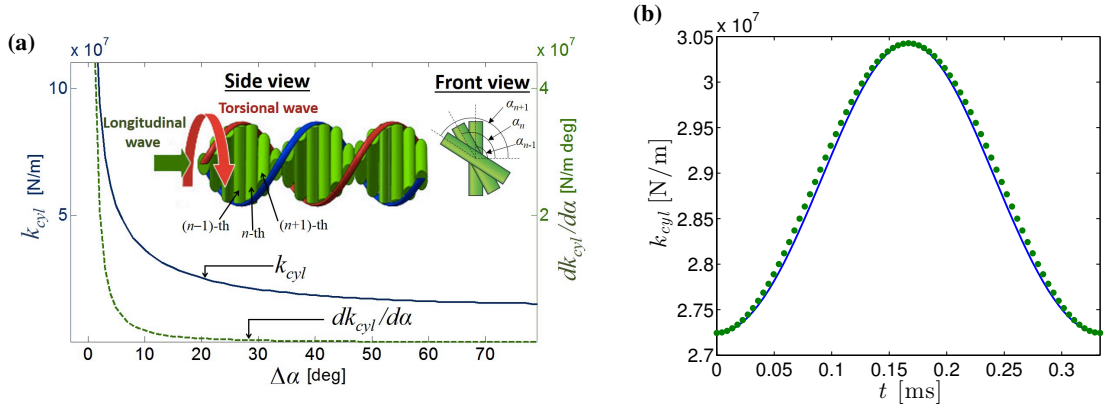

    \centering
  \begin{tabular}{@{}p{0.5\linewidth}@{\quad}p{0.44\linewidth}@{}}
    \subfigimg[width=\linewidth]{\bf (a)}{stiffness1newb.pdf} &
    \subfigimg[width=\linewidth]{\bf (b)}{K-taylor-eps-converted-to.pdf} 
  \end{tabular}
 \caption{\textbf{(a)} Stiffness and sensitivity of the cylindrical Hertzian contact as a function of alignment
angles. 
A schematic of the HPC composed of a 1D chain of helicoidally stacked cylindrical
particles under precompression is shown in the inset. \textbf{(b)} The stiffness coefficient $k_{cyl}(\Delta \alpha)$ (solid blue line) and its single harmonic approximation based on its Taylor expansion (green markers). The parameter values in this case are
 $\alpha_0 = 10^\circ $, $A_\alpha=1^\circ$ and $f_\alpha = 3$ kHz.  }
 \label{fig:stiffness}
\end{figure}

\begin{figure}
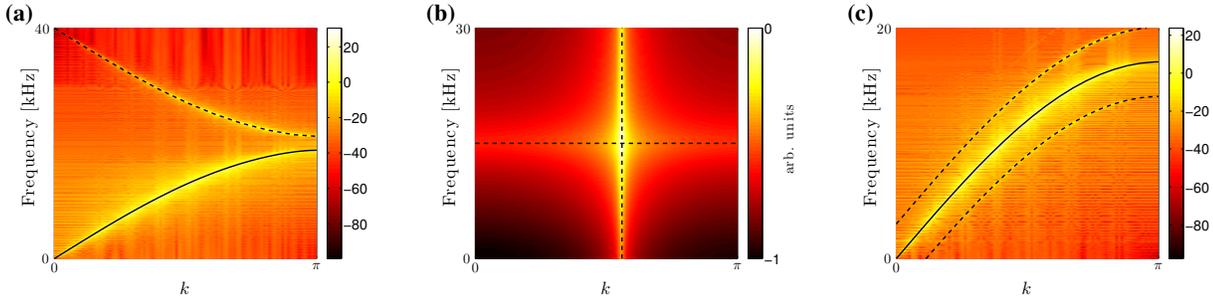

  \centering
  \begin{tabular}{@{}p{0.33\linewidth}@{\quad}p{0.33\linewidth}@{\quad}p{0.33\linewidth}@{}}
    \subfigimg[width=\linewidth]{\bf (a)}{stable_time-eps-converted-to.pdf} &
     \subfigimg[width=\linewidth]{\bf (b)}{unstable_time-eps-converted-to.pdf} & 
    \subfigimg[width=\linewidth]{\bf (c)}{semistable_time-eps-converted-to.pdf} 
  \end{tabular}
\caption{\textbf{(a)}: Superposition of: {(i)} Floquet exponents $\sigma(k)/(2\pi)$ (solid line) and a first order harmonic $\sigma(k)/2\pi + f_\alpha $ (dashed line) versus the wavenumber $k$
 (parameter values are  $\alpha_0=10$, $A_\alpha=5$ and $f_\alpha = 40$ kHz which satisfy the stability condition~\eqref{eq:stability})
and  {(ii)} 
 PSD [dB] of velocity component of the solution of Eq.~\eqref{eq:eomNL} obtained by imposing a chirped pulse (0 to 30 kHz) with an amplitude of $5$ nm on the left end of a resting chain. 
 \textbf{(b)}:  PSD [arbitary units] of velocity component of the solution with the unstable parameter values
  $\alpha_0=10$, $A_\alpha=5$ and $f_\alpha = 30$ kHz. The vertical dashed line corresponds to the wavenumber yielding the Floquet exponent
 with the largest real part. The horizontal line is the corresponding imaginary part of that exponent. 
 The chain was excited in the same way as in panel (a).
 \textbf{(c)}: Same as panel (a), but with the parameter values  $\alpha_0=10$, $A_\alpha=1$ and $f_\alpha = 3$ kHz, which do not satisfy the stability condition~\eqref{eq:stability}. However, in this case, the wavenumbers do not fall into regions of instability of the Mathieu
 equation due to the finite nature of the simulations (see text), and thus the observed dynamics are stable. The chain was excited in the same way as in panel (a).
  }
 \label{fig:disp_time}
\end{figure}

We now consider a standing torsional wave in the helicoidal chain, where the angles of the
cylindrical particles are given by:
\begin{equation}\label{eq:talpha}
 \alpha_n = n\alpha_0 + n \, A_\alpha \cos(\omega_\alpha t)  
\end{equation}
where $\omega_\alpha = 2 \pi f_\alpha$ is the frequency of the standing torsion wave. These angular variations result in dynamic
stiffness changes in the axial direction.
In order to simplify the analysis, we assume $|A_\alpha| \ll |\alpha_0|$
such that we may Taylor expand $k_{cyl}(\Delta \alpha)$ with respect to $A_\alpha \cos(\omega_\alpha t)$,
yielding an expression for the linear stiffness that is composed of a single harmonic
\begin{equation} \label{eq:ktime}
 k_{lin} \approx \tilde{\alpha}_0 + \tilde{A}_\alpha \cos(\omega_\alpha t)  
\end{equation}
where
$$\tilde{\alpha}_0=  \frac{3}{2} k_{cyl}(\alpha_0) \delta_0^{1/2}        \qquad  \tilde{A}_\alpha = \frac{3}{2} k'_{cyl}(\alpha_0) A_{\alpha} \delta_0^{1/2},   $$
where the prime denotes the derivative with respect to the argument.
See e.g. Fig.~\ref{fig:stiffness}(b) for the validity of this
approximation. In order to compute the dispersion relationship of~\eqref{eq:eom} with the time-dependent
stiffness coefficient~\eqref{eq:ktime} we make use of the discrete Fourier transform
$$\hat{u}(k,t) = \sum_{n\in\Z} u_n(t) e^{ikn} $$ 
%
where $\hat{u}(k,t) = \hat{u}(k+\pi,t)$.
Thus, Eq.~\eqref{eq:eom} can be rewritten as
\begin{equation}\label{eq:Mat}
 \pa_t^2\hat{u}(k,t) = -\omega(k)^2 \left( 1 +  \frac{\tilde{A}_\alpha}{\tilde{\alpha}_0} \cos(\omega_\alpha t) \right) \hat{u}(k,t) 
\end{equation}
where $\omega(k)$ is the dispersion relation in the homogeneous case, see Eq.~\eqref{eq:disp_hom}. For each $k$, Eq.~\eqref{eq:Mat}
represents the well-studied Mathieu equation, whose general solution can be found using Floquet theory \cite{Mathieu}:
$$ \hat{u}(k,t) = c_1 e^{\mu_1(k) t} p_1(t) + c_2 e^{\mu_2(k) t} p_2(t) $$
where $\mu_1(k)$ and $\mu_2(k)$ are the Floquet exponents, $c_1$ and $c_2$ are arbitrary constants, and $p_1$ and $p_2$ are functions with period $T_\alpha = 2\pi/\omega_\alpha$. 
For the Mathieu equation, stability is only possible if the Floquet exponents are purely imaginary. Typically, the imaginary part of the exponent (which we denote $\sigma(k)$) is incommensurate
with the frequency $\omega_\alpha$, and thus the dynamics are not periodic. The unstable and stable regions are separated in parameter space by periodic solutions.
Thus, the stability boundaries can be determined by finding parameter values where periodic solutions are possible. This can be achieved by substituting 
the Fourier series representation of a periodic solution into the Mathieu equation, and demanding that the Fourier coefficients be non-trivial.  Doing so leads
to the well known determinant condition to determine the stability regions of the Mathieu equation \cite{Wei}.
In the $ \displaystyle \left( \frac{\omega_{\alpha}}{2 \omega(k)}, \frac{\tilde{A}}{\tilde{\alpha}_0} \right) $ plane, the regions of instability
are wedge like and originate (i.e. when $\tilde{A}/\tilde{\alpha}_0=0$) at the values
\begin{equation} \label{eq:wedge}
\frac{\omega_\alpha}{2 \omega(k)} = \frac{1}{j}
\end{equation}
where $j\in\Z^+$. 
Since in our setting $|\omega(k)|$ increases monotonically from zero as $k$ increases, we need only to consider the first stability boundary (i.e. the one originating at  $\frac{\omega_\alpha}{2 \omega(k)} = 1$). This boundary can be approximated analytically, leading to following
condition for stability in our setting:
\begin{equation}\label{eq:stability}
  \frac{\omega_\alpha}{2 \omega(\pi)} > 1 + \left| \frac{\tilde{A}}{4 \tilde{\alpha}_0} \right|.  
\end{equation}
Figure~\ref{fig:disp_time}(a-b) shows the Fourier transform in the spatial and temporal domains of a solution of Eq.~\eqref{eq:eomNL}
for parameter values satisfying and violating condition~\eqref{eq:stability}, respectively. 
For the unstable case considered in Fig.~\ref{fig:disp_time}(b), only the first instability region is entered.
Thus, although the solutions grow without bound, one can detect
spectral concentration about wavenumbers that fall in the instability region.
For a chain of finite length $L$, the span of wavenumbers becomes discrete, e.g.
$k_n = n \pi / (L+1)$ in the case of fixed boundary conditions. Thus,  it is possible that each wavenumber avoids each instability region of the Mathieu equation. For example,
if we assume fixed boundary conditions and $|\tilde{A}|  \ll |\tilde{\alpha}_0| $ then Eq.~\eqref{eq:wedge} implies 
that the quantity 
\begin{equation} \label{eq:stability_finite}
\frac{\omega_\alpha}{2 \omega(n \pi / (L+1))}, \qquad |n| \leq L 
\end{equation}
should not be in a neighborhood of $1/j$ for each $j\in\Z^+$ in order to achieve stability. 
See  Fig.~\ref{fig:disp_time}(c) for an example.

In the case
that the Mathieu equation is stable, we have $\mu_{1,2}(k) = \pm \, i \sigma(k)$, where $\sigma(k)\in\R$ and $p_1(t) = \bar{p}_2(t)$.
Since $p_1$ has period $T_\alpha$, we have the following solution
\begin{equation}\label{eq:sol}
 \hat{u}(k,t) =  \sum_{m\in \Z} a_m(k) e^{i(\sigma(k) + \omega_\alpha m) t}, \qquad a_m \in \C
\end{equation}
%
%
where the Fourier coefficients $a_m(k)$ of the function $p_1(t)$ satisfy
%
\begin{equation}\label{eq:fc}
    \frac{a_m(k)}{a_{m-1}(k) + a_{m+1}(k)} = \frac{\tilde{A}_\alpha}{2 \tilde{\alpha}_0} \left( \frac{\omega(k)^2}{ (\sigma(k)+\omega_\alpha m)^2 - \omega(k)^2  } \right).
\end{equation}
Our analysis shows that the longitudinal wave, interacting with the periodic stiffness variations, produces
frequency shifts similar to optical Brillouin scattering~\cite{brillouin}. This
implies that the HPCs can realize wave mixing effects, whereby interactions between
longitudinal and torsional waves produce extra dispersion modes in variable frequencies.
Thus, the newly created modes in time-variant HPCs effectively up-shift cutoff
frequencies ($f_{\rm cutoff} \rightarrow f_{\rm cutoff} + f_\alpha $), 
as is shown in e.g. Fig.~\ref{fig:disp_time}(a,b). This
is especially useful for application purposes, a theme
that we will revisit in Sec.~\ref{sec:transistor}.
It is also relevant to point out that  $a_{\pm 1}$ associated
with $\sigma(k) \pm \omega_\alpha$ already bear an amplitude nearly two orders
of magnitude smaller than the principal mode $a_0$. Hence, it is natural
to expect that $a_{\pm 2}$ are considerably harder to identify in the context
e.g. of Fig.~\ref{fig:disp_time}(a,b).

As an additional comment, we should note that here we consider angular
periodic variations in the cosinusoidal form, in order to
capture the essential characteristics of the impact of this variation
(in the spirit also of Fourier decomposition of any periodic variation
into such modes). 
Moreover, the evolution of the modes in this case are 
described by the Mathieu equation~\cite{Wei} which is well-established in the realm of
parametric instabilities. 
Nevertheless, it would be worthwhile to consider
other, special forms of periodic variation within the more
broad setup of Hill's equations~\cite{hill} as generalizations of
Eq.~(\ref{eq:Mat}). A notable example that is
worthy to explore from the point of view of deriving explicit 
analytical conditions for the instability manifestation is that
of piecewise constant variations of the angle, along the lines
of the well-known, exactly
solvable Kronig-Penney model of quantum mechanics~\cite{goldman}.

\vspace{2cm}

\section{Space-variant, time-independent HPCs} \label{sec:space}

\begin{figure}[t]
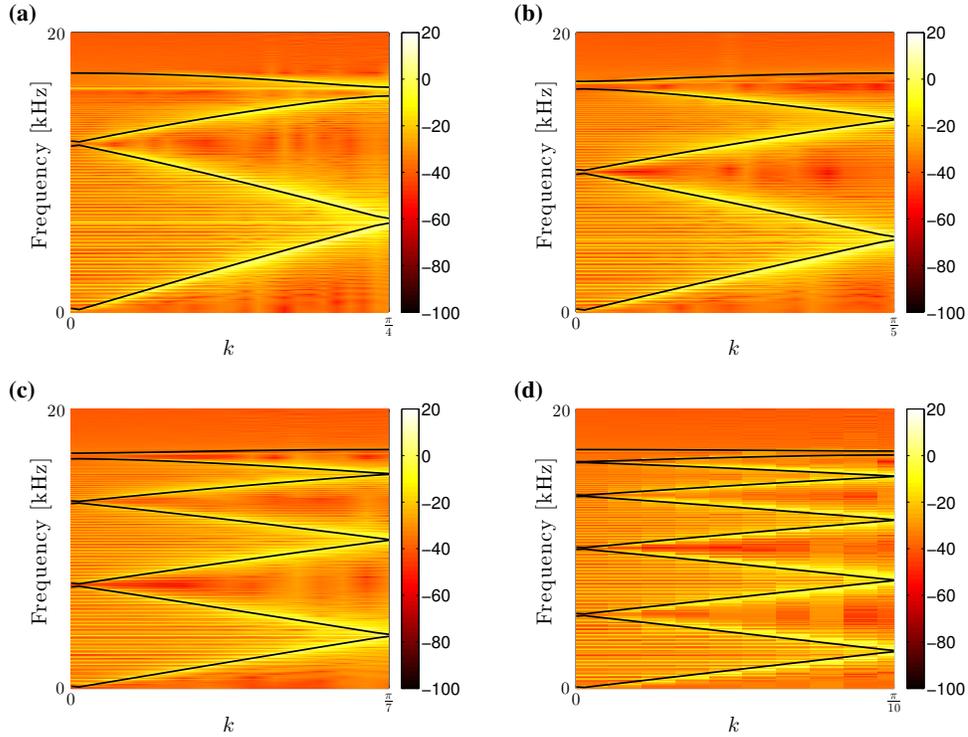

    \centering
  \begin{tabular}{@{}p{0.4\linewidth}@{\quad}p{0.4\linewidth}@{}}
    \subfigimg[width=\linewidth]{\bf (a)}{N3-eps-converted-to.pdf} &
    \subfigimg[width=\linewidth]{\bf (b)}{N5-eps-converted-to.pdf} \\
    \subfigimg[width=\linewidth]{\bf (c)}{N7-eps-converted-to.pdf} &
    \subfigimg[width=\linewidth]{\bf (d)}{N10-eps-converted-to.pdf} 
  \end{tabular}
 \caption{\textbf{(a)} Dispersion relation in the case of space-dependent 
torsional waves. Shown is the superposition of 
(i) $N=3$ dispersion curves (solid lines) versus the wavenumber $k\in [0,\pi/N]$
 (parameter values are $\alpha_0 = 10^{\circ}$ and $A_\alpha = 1^{\circ}$)
and  (ii) 
 PSD [dB] of velocity component of the solution of Eq.~\eqref{eq:eomNL} obtained by imposing a chirped pulse (0 to 30 kHz) with an amplitude of $5$ nm on the left end of a resting chain. 
 \textbf{(b)-(d)} Same as (a) but with  (b) $N=5$, (c) $N=7$ and (d) $N=10$.}
 \label{fig:disp_space}
\end{figure}

We now consider modulating the rotational angles of the helicoidal chain in the spatial domain.
Under, once again, harmonic perturbations but now in space, 
the angles of the cylindrical particles become:
$$ \alpha_n = n \alpha_0 + A_\alpha \cos(k_{\alpha} n )  $$
where $A_\alpha$ and $k_\alpha$ are the modulation amplitude and the wavenumber ($k_\alpha = 2\pi / N$, where $N$ is the
spatial period of angular variations). 
In this case, the linear stiffness coefficient satisfies
$$ k_{lin}(\Delta \alpha_n) = k_{lin}(\Delta \alpha_{n+N}).$$
Thus, rather than use the Fourier transform to compute the dispersion
relationship (as in the case of spatially homogeneous media), we use the Bloch transform \cite{Hussein} 
\begin{equation}\label{eq:Bloch_trans}
\check{u}_j(k,t) =  \sum_{n\in\Z} u_{nN +j}(t)e^{iknN}  
\end{equation}
where $ j \in \{1,2,\ldots,N\}$ and $\check{u}_j(k,t) = \check{u}_j(k + 2\pi/N,t)$. Applying the Bloch transform to~\eqref{eq:eom} yields
\begin{equation}\label{eq:Bloch}
M \pa_t^2\check{u}_j(k,t) = k_{lin}(\Delta \alpha_j)\check{u}_{j-1}(k,t) - (k_{lin}(\Delta \alpha_j)+ k_{lin}(\Delta \alpha_{j+1}))\check{u}_j(k,t) + k_{lin}(\Delta \alpha_{j+1})\check{u}_{j+1}(k,t).
\end{equation}
This system of $N$ equations has solutions of the form $ \check{u}_j(k,t) = \exp(ijk + i\omega_B(k) t) f_j(k)$ where
${\bf f}$ and $\omega_B$ satisfy the eigenvalue problem

\begin{equation}\label{EVproblem}
- \omega_B^2(k) {\bf f}(k)   = M^{-1} B(k) {\bf f}(k) 
\end{equation}
where $ {\bf f} = (f_1,f_2,\ldots,f_N)^T $, and 
\begin{equation} \label{eq:B}
B(k) = \left( \begin{array}{cccccccc} b_{1,1} & b_{1,2} & 0 &\hdots & 0 &0 & b_{1,N} \\
b_{2,1} & b_{2,2} & b_{2,3} & 0 & \hdots & 0 & 0 \\ 
0 & b_{3,2} & b_{3,3} & b_{3,4} & 0 &\hdots & 0 \\
\\ \vdots &&\ddots &\ddots &\ddots&& \vdots  \\ \\ 
0 &\hdots &0 & b_{N-3,N-2} & b_{N-2,N-2} & b_{N-2,N-1} & 0 \\
0 & 0 & \hdots &0 & b_{N-2,N-1} & b_{N-1,N-1} & b_{N-1,N} &  \\
b_{N,1} & 0 &0 & \hdots&0 & b_{N,N-1} & b_{N,N} 
 \end{array} \right) 
\end{equation}
with 
\begin{eqnarray*}
b_{j,j+1} & = &   k_{lin}(\Delta \alpha_{j+1})  e^{i k}, \\
b_{j,j} & = &   -(k_{lin}(\Delta \alpha_{j}) + k_{lin}(\Delta \alpha_{j+1})) , \\
b_{j,j-1} & = &  k_{lin}(\Delta \alpha_{j})   e^{-i k},\\
b_{1,N} & = &  k_{lin}(\Delta \alpha_{1})   e^{-i k},\\
b_{N,1} & = &  k_{lin}(\Delta \alpha_{1})  e^{i k}
\end{eqnarray*}
for $N>2$, and 
\begin{eqnarray*}
b_{1,2} & = &   k_{lin}(\Delta \alpha_{2}) e^{i k} + k_{lin}(\Delta \alpha_{1})e^{-i k} ,   \\
b_{j,j} & = &   -(k_{lin}(\Delta \alpha_{j}) + k_{lin}(\Delta \alpha_{j+1})) , \\
b_{2,1} & = &  k_{lin}(\Delta \alpha_{1}) e^{i k} + k_{lin}(\Delta \alpha_{2})e^{-i k} 
\end{eqnarray*}
for $N=2$. If the system parameters are chosen such that $k_{lin}(\Delta \alpha_{j})  >0$, then stability follows trivially, since $B$ has real, non-positive eigenvalues (which can be shown using the Gershgorin circle theorem~\cite{atkinson} or by noting that $B$ is a Jacobi operator \cite{Teschl}). Under this assumption,
our dispersion relation will be composed of $2N$ curves ($N$ of which are non-negative) see e.g. Fig.~\ref{fig:disp_space}(a). 
We see that multiple wave modes are generated due to the
effect of periodic variations of spatial angles and that band gaps appear among newly generated dispersion curves.
It should be noted that in other areas of physics, such as e.g.
nonlinear optics~\cite{gligoric} and atomic Bose-Einstein 
condensates~\cite{pearl_pra} (see also references therein), the
use of such so-called superlattice potentials is fairly
widespread, leading to the formation of mini-gaps (i.e.,
gaps within the fundamental band existence in the absence
of additional 
periodicities). Within these mini-gaps, nonlinear stationary
states are also typically sought in these problems.

The space-time evolution of the longitudinal waves can be obtained from the direct
numerical integration of Eq~\eqref{eq:eomNL}. Again, we impose a chirped pulse (0 Hz to 30 kHz) on one end of
the chain with amplitude of 5 nm. The space-time evolution of the HPC's dispersion behavior is
calculated via the Fourier transform in the temporal domain and the Bloch transform~\eqref{eq:Bloch_trans} in the spatial domain
of the bead's velocities, see Fig.~\ref{fig:disp_space}. While these  mini-gaps become 
possible for space-dependent torsional waves, we note 
they become smaller as $N$ increases, see Fig.~\ref{fig:disp_space}(b-d).
As the Brillouin zone shrinks to accommodate the larger periodicity
index $N$ (hence the progressively narrower zone ending at $\pi/N$),
$N$ segments of the dispersion relation ``fold'' inside this narrower
zone with the progressively also narrower mini-gaps separating them
as $N$ increases. This trend 
is clearly illustrated in Fig.~\ref{fig:disp_space}.

%

\section{Space- and Time-Varying HPCs}\label{sec:spacetime}

\begin{figure}
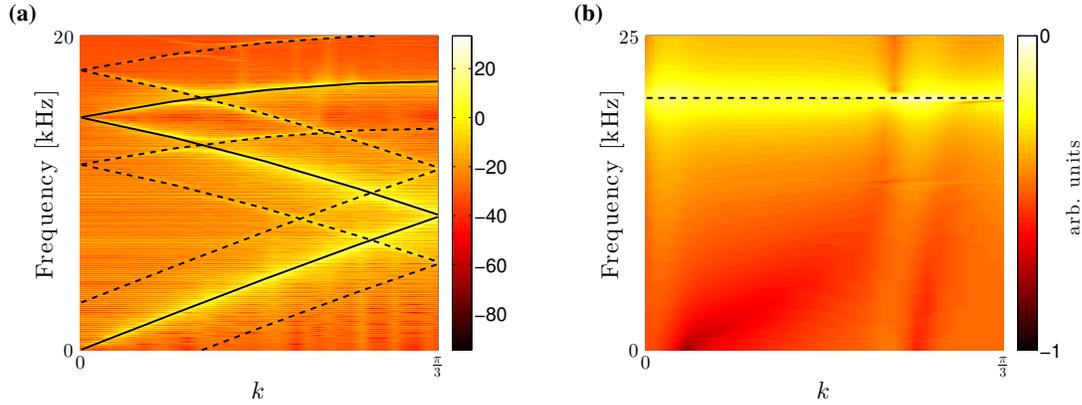

     \centering
  \begin{tabular}{@{}p{0.45\linewidth}@{\quad}p{0.45\linewidth}@{}}
    \subfigimg[width=\linewidth]{\bf (a)}{semistable_spacetime-eps-converted-to.pdf} &
    \subfigimg[width=\linewidth]{\bf (b)}{unstable_spacetime-eps-converted-to.pdf} 
  \end{tabular}
 \caption{\textbf{(a)} Dispersion relation in the case of space- and time-dependent torsional waves
 for $\alpha_0 = 10^{\circ}$, $A_\alpha = 2^{\circ}$, $N=3$ and $f_\alpha = 3$ kHz.  Shown is
 the superposition of  (i)  $N=3$ Floquet exponents (solid lines) and the first
 order harmonic shifts $\sigma(k)/2\pi \pm f_\alpha$ (dashed lines),
and (ii) 
 PSD [dB] of velocity component of the solution of Eq.~\eqref{eq:eomNL} obtained by imposing a chirped pulse (0 to 30 kHz) with an amplitude of $5$ nm on the left end of a resting chain.
 \textbf{(b)}  PSD [arbitrary units] of velocity component of the solution with unstable parameter values
  $\alpha_0=10$, $A_\alpha=5$, $N=3$ and $f_\alpha = 40$ kHz (note this parameter set is stable for $N=1$, see e.g. Fig.~\ref{fig:disp_time}(a)).  In this case, the Floquet multiplier with the largest modulus occurs for $k\approx 0$. The horizontal line is the imaginary part of the corresponding Floquet exponent. 
 The chain was excited in the same way as in panel (a).
 }
 \label{fig:disp_timespace}
\end{figure}

Here we combine both space- and time-variant effects in the form of traveling torsional
waves:
$$ \alpha_n = n \alpha_0 + A_\alpha \cos(k_{\alpha} n  + \omega_\alpha t)  $$
where $A_\alpha$ is the modulation amplitude,  $k_\alpha$ is the wavenumber ($k_\alpha = 2\pi / N$, where $N$ is the
spatial period of angular variations) and  $\omega_\alpha = 2\pi f_\alpha$ is the frequency of the traveling torsion wave.
Thus, the linear stiffness coefficient now satisfies
$$ k_{lin}(\Delta \alpha_n)(t) = k_{lin}(\Delta \alpha_{n+N})(t),  \qquad k_{lin}(\Delta \alpha_n)(t) = k_{lin}(\Delta \alpha_{n})(t + 2\pi/\omega_\alpha)   $$
Applying the Bloch transform~\eqref{eq:Bloch_trans} to \eqref{eq:eom} in this case leads to the system of $N$ second order ODEs with time-periodic coefficients
\begin{equation} \label{eq:msystem}
M \pa_t^2 \check{\bf u}(k,t) = B(k,t) \check{\bf u}(k,t)
\end{equation}
where $\check{\bf u} = (\check{u}_1,\check{u}_2,\ldots,\check{u}_N)^T$ and $B$ is defined by~\eqref{eq:B} (note in this case  $B(k,t) = B(k,t + 2 \pi/ \omega_\alpha) )$.   
Unlike like the time- varying set-up considered in Sec.~\ref{sec:time}, there are no analytically tractable conditions for parametric stability of Eq.~\eqref{eq:msystem}. Thus, one needs to 
numerically compute the $2N$ Floquet multipliers to verify that none has modulus greater
than unity (where the multiplier is defined as $\exp(\mu T_\alpha)$). 
Assuming stability, the dispersion curves will have the form 
$\sigma_j(k) + m \omega_\alpha $, where $\sigma_j(k)$ is the imaginary part of the $j$th Floquet exponent and $m\in\Z$.  Motivated by Fig.~\ref{fig:disp_time}(b), we choose parameter values with
$|A_{\alpha}|  \ll |\alpha_0|$, which leads to a stable system,
see e.g. Fig.~\ref{fig:disp_timespace}(a). In this case, as expected, the spectrum is altered by applying an up-shift to the cutoff (due to the
temporal variance, like in Sec.~\ref{sec:time}) and by increasing the number of primary ($m=0th$ order) branches (due to the
spatial variance, like in Sec.~\ref{sec:space}). 
Essentially, in this case, we see a combination of
the phenomenologies of Secs. \ref{sec:time} and \ref{sec:space}.
The latter is responsible for the formation of the mini-gaps and
the folding of the dispersion relation within the narrower Brillouin
zone while the former is responsible for the emergence of, chiefly,
the shifted frequencies by $\pm m  \omega_\alpha$ (once again, we 
chiefly observe the ones with $m=1$). An interesting observation in this case, is the apparent avoided crossing
of the Floquet exponents of one branch, with the higher order harmonics of another branch, as seen in Fig.~\ref{fig:disp_timespace}(a).
This is a well-known feature of  (typically self-adjoint) matrices
representing physical systems under mono-parametric
variations~\cite{lax}. It has to do with the fact that degenerate
matrices with multiple eigenvalues form a surface of codimension 2
and hence such a crossing cannot be typically created by a mono-parametric
tuning, such as the one considered e.g. in Fig.~\ref{fig:disp_timespace}(a).

It is also worth noting that the presence of spatial invariance can destabilize the system.
For example, in Fig.~\ref{fig:disp_timespace}(b) we consider the same stable parameter
values as in Fig.~\ref{fig:disp_time}(a), but with a spatial period of $N=3$, leading to
instability.

\section{An Application Based on HPCs: Towards an Acoustic Transistor} \label{sec:transistor}

 \begin{figure}
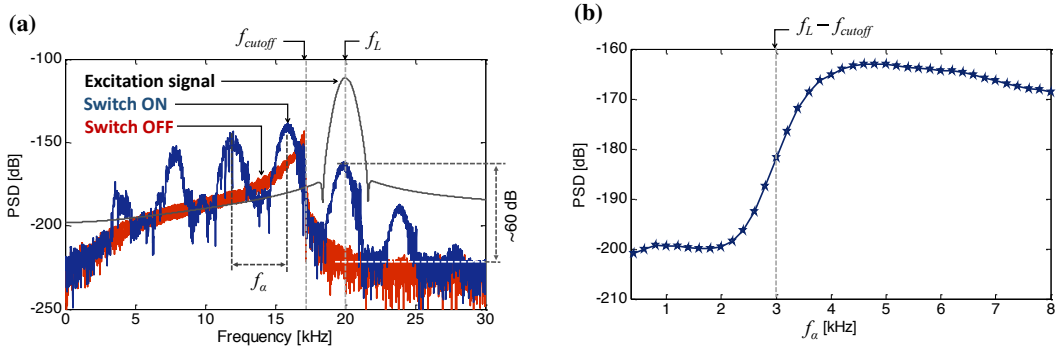

    \centering
  \begin{tabular}{@{}p{0.45\linewidth}@{\quad}p{0.45\linewidth}@{}}
    \subfigimg[width=\linewidth]{\bf (a)}{Transistor1.pdf} &
    \subfigimg[width=\linewidth]{\bf (b)}{Transistor2.pdf} 
  \end{tabular}
 \caption{\textbf{(a)} The transistor effect of HPCs. The black curve denotes the Gaussian input signal, while the red (blue) curve corresponds to transmitted signals under the switch off (on) state. $f_L$ and $f_{\rm cutoff}$ are input and cutoff frequencies. \textbf{(b)} Variation of the transmission of the acoustic wave at frequency $f_L$ as a function of the gate frequency $f_\alpha$.  }
 \label{fig:transistor}
\end{figure}

The realm of HPCs through its tunability 
can offer a platform that may facilitate 
the realization of devices analogous to well-established electronic
ones. Here, we mention an example of this type in the form of an 
acoustic transistor.
The HPCs can lead to features similar to those of the transistor
based on the finding that the dispersive bands of axial waves can be 
shifted actively by torsional waves. To test this, we numerically excite 
the first particle in the chain with a Gaussian pulse and generate longitudinal waves, whose central frequency $f_L$ 
is chosen to be above the cutoff frequency $f_{\rm cutoff}$. See Fig.~\ref{fig:transistor}(a) for example with the parameter values $f_L = 20$ kHz, $b_w = 1$ kHz, $f_{\rm cutoff} = 17.0$ kHz,
$A_\alpha = 1^\circ, f_\alpha  = 4$ kHz and $N=10$. The PSD of the input is the blue curve and the transmitted wave measured at the end of the HPC is the red curve. We find that the longitudinal wave above the cutoff frequency is blocked by the chain in an evanescent manner, and the transmission gain is about $-220$ dB. This corresponds to an off state. Now we apply a torsional standing wave and we observe that the amplitude of the longitudinal wave increases by 60 dB (blue curve in Fig.~\ref{fig:transistor}(a)). This is the on state, which confirms the efficiency of HPCs towards tuning longitudinal propagation,
by means of the applied torsional standing wave. Additionally, the transmitted signals contain harmonics of $f_L$ and $f_\alpha$ such as $f_L - f_\alpha (16.0 kHz)$ and  $f_L + f_\alpha (24.0 kHz)$. It is important to note that
while this type of signal control is strongly reminiscent of the
functionality of a(n acoustic) transistor, nevertheless, our setup
does not possess the amplification characteristics encountered
in a regular transistor and hence our ``device'' should not be
considered an acoustic transistor per se.
    
From the working principle of the above setup, the following condition should be satisfied for the transmission of longitudinal waves: 
\begin{equation}\label{eq:trans}
 f_\alpha > f_L - f_{\rm cutoff}.
 \end{equation}
Thus, there should exist a lower frequency threshold for a given $f_L$ and $f_{\rm cutoff}$. In order to show this threshold effect, we investigate the frequency dependence of the transmission. We excite a Gaussian pulse with the center frequency $f_L = 20.0$ kHz and bandwidth $b_w = 120$ Hz. 
The numerical results of the switch transmission with respect to $f_\alpha$ are shown in Fig.~\ref{fig:transistor}(b). The lower frequency threshold, $f_L - f_\alpha$, according to Eq.~\eqref{eq:trans} is plotted by the vertical line. We can see that the transmission of longitudinal wave increases significantly at $f_L - f_\alpha$. This feature is also reminiscent of an electrical 
transistor: when the gate source voltage is higher than threshold 
voltage, the conducting channel begins to connect the source and drain 
of the transistor, allowing a large current to flow.


\section{Conclusions and Future Challenges} \label{sec:theend}

We investigated the characteristics of helicoidal phononic crystals
(HPCs) in a shape similar to DNA architectures. Based on the 
Hertzian contact
among slanted cylindrical elements, the HPCs develop strong cross-talking between in-plane
torsional waves and out-of-plane longitudinal waves. Our (semi)-analytical dispersion computations demonstrated that the HPCs
exhibit versatile, controllable behavior of longitudinal wave transmission as a function of spatial and
temporal-variations of torsional waves, which was confirmed against full numerical simulations of the pertinent model.
Specifically, it was shown that time-variant
HPCs show (symmetric up- and down-) shifts of dispersive wave modes. 
This was used to demonstrate that the longitudinal waves can be switched on/off by torsional waves, in an effect reminiscent of the electrical
transistor. 
On the other hand, the space-dependent variant was found to appropriately
modify the dispersion relation introducing a reduced Brillouin zone
and corresponding mini-gaps within the linear spectrum. The combination
of the two effects provided a combination of their influences, as well
as additional intriguing features, such as the observed avoided
level crossings.
Our conclusion is that
the time- and space-dependent phononic crystal provides an 
ideal setting to manipulate
acoustic waves by leveraging wave mixing and switching effects and opens the doors for a host of other studies.

Among the themes of immediate interest,  we include the effects of coupling the dynamics of the torsional and longitudinal waves, and considering higher amplitude
excitations to test what benefits/disadvantages the nonlinearity of the system introduces. In the latter setting exploring the periodicity influence
in the context of traveling waves~\cite{Nester01,Sen}, and especially of 
the 
localization effects towards the formation of more complex localized
breather~\cite{dar01} excitations.
It should be noted here that these settings are not only amenable
to direct theoretical/numerical investigations, but additionally
we believe should within the immediate grasp of current, state-of-the-art
experiments in the field. 

\section*{Acknowledgement}
We thank the support from the US-AFOSR (FA9550-12-1-0332), 
ONR (N000141410388), and NSF (CMMI-1234452,
CMMI-1000337, CMMI-844540). We are grateful to G. Gantzounis for his valuable 
input on this work.

\appendix

\section{Decay of Fourier modes in time-variant, space-independent HPCs}

The presence of the first order harmonic shifts (see e.g. Fig.~\ref{fig:disp_time}(b))
is what ultimately allows a signal with frequency content lying outside the passband to be
transmitted through the chain. Thus, to improve the efficiency of this ``transistor effect", it would be desirable to have control over the amplitude of the first order harmonics relative to the amplitude of the original dispersion curve. This can be achieved by assuming $|a_m| \ll 1$ for $|m|>1$ in Eq.~\eqref{eq:fc}, yielding the following prediction
\begin{eqnarray}
    \frac{a_1(k)}{ a_{0}(k)} &=& \frac{\tilde{A}_\alpha}{2 \tilde{\alpha}_0} \left( \frac{\omega(k)^2}{ (\sigma(k) + \omega_\alpha )^2 - \omega(k)^2  } \right),  \label{eq:decay1} \\
        \frac{a_{-1}(k)}{ a_{0}(k)} &=& \frac{\tilde{A}_\alpha}{2 \tilde{\alpha}_0} \left( \frac{\omega(k)^2}{ (\sigma(k) - \omega_\alpha )^2 - \omega(k)^2  } \right).  \label{eq:decay2} 
\end{eqnarray}

\begin{figure}[t]
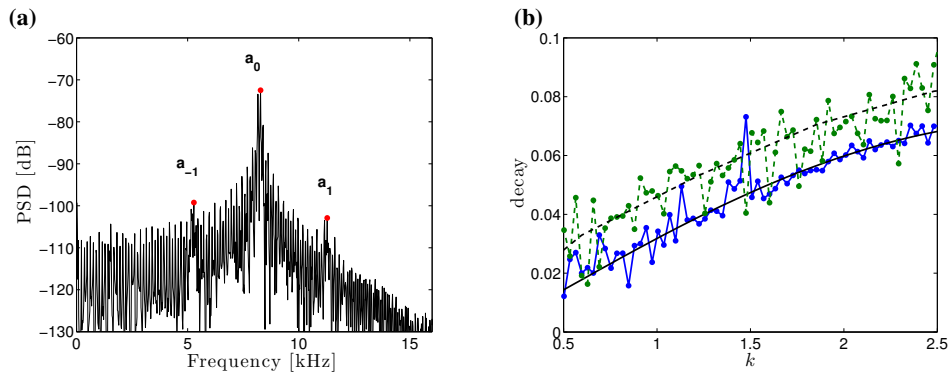

   \centering
  \begin{tabular}{@{}p{0.4\linewidth}@{\quad}p{0.4\linewidth}@{}}
    \subfigimg[width=\linewidth]{\bf (a)}{decay_profile-eps-converted-to.pdf} &
    \subfigimg[width=\linewidth]{\bf (b)}{decay_k-eps-converted-to.pdf} 
  \end{tabular}
 \caption{\textbf{(a)} PSD of the displacements $u_n$ of the solution shown in Fig.~\ref{fig:disp_time}(b) for the wavenumber $k=1$. The spectral peaks corresponding to $a_{-1},a_0$ and $a_{1}$ are shown as red points. \textbf{(b)}
 The decay  $a_{1}/a_0$ (green dashed line with markers) and $a_{-1}/a_0$ (blue line with markers) versus the wavenumber $k$. The analytical predictions from Eq.~\eqref{eq:decay1} (black dashed line) and Eq.~\eqref{eq:decay2} (solid black line) are also shown.}
 \label{fig:decay}
\end{figure}

To verify this prediction, we numerically solve Eq.~\eqref{eq:eomNL} and compute the Fourier transform in the time domain for each wavenumber $k$.  The largest
peak corresponds to the $a_0$ mode, which is concentrated around the frequency $\sigma(k)$ and the modes $a_{1}$ and $a_{-1}$ are concentrated
at $\sigma(k) \pm \omega_\alpha$, see Fig.~\ref{fig:decay}(a) for example. In Fig.~\ref{fig:decay}(b) the decay prediction
is shown against numerically computed values for various wavenumbers $k$, where the trend is captured well. Discrepancies
stem from the finite nature of the domains considered and the approximation $|a_m|=0$ for $|m|>1$. From Eq.~\eqref{eq:decay1}-\eqref{eq:decay2}, we see that if one desires larger spectral peaks at the $m=\pm 1$ harmonics, larger values of $|\tilde{A}_\alpha / \tilde{\alpha}_0|$  must be taken. However,
this must be weighed against the stability condition~\eqref{eq:stability} where large values of $| \tilde{A}_\alpha/\tilde{\alpha}_0 |$ can cause instabilities.

\begin{thebibliography}{99}

\bibitem{Sanchez} J. V. S\'anchez-P\'erez, D. Caballero, R. M\'artinez-Sala, C. Rubio, J. S\'anchez-Dehesa, F. Meseguer, J. Llinares, and F. G\'alvez PRL \textbf{80},  5325 (1998).
\bibitem{Liu} Z. Y. Liu,   Xixiang Zhang, Yiwei Mao, Y. Y. Zhu, Zhiyu Yang, C. T. Chan, Ping Sheng, Science \textbf{289}, 1734 (2000).
\bibitem{Khelif} A. Khelif,   P. A. Deymier, B. Djafari-Rouhani, J. O. Vasseur and L. Dobrzynski, Journal of Applied Physics \textbf{94}, 1308 (2003).
\bibitem{Lin} S.-C. S. Lin, and T. J. Huang, Physical Review B \textbf{83}, 174303 (2011).
\bibitem{Jun} Jun Liu, Yihui Wu, Feng Li, Ping Zhang, Yongshun Liu and Junfeng Wu, EPL (Europhysics Letters) \textbf{98}, 36001 (2012).
\bibitem{Wright} D. W. Wright, and R. S. C. Cobbold, Smart Materials and Structures \textbf{18}, 015008 (2009).
\bibitem{Narisetti} R. K. Narisetti, M. Ruzzene, and M. J. Leamy, Wave Motion \textbf{49}, 394 (2012).
\bibitem{Yang} J. Yang, S. Dunatunga, and C. Daraio, Acta Mechanica \textbf{223}, 549 (2012).
\bibitem{bertoldi} P. Wang, F. Casadei, S. Shan, J.C. Weaver,
K. Bertoldi, 
Phys. Rev. Lett. {\bf 113}, 014301 (2014).
\bibitem{Nester01} V. F. Nesterenko, 
{\it Dynamics of Heterogeneous Materials}, Springer-Verlag
(New York, 2001).
\bibitem{Sen} Surajit Sen, Jongbae Hong, Jonghun Bang, Edgar Avalos, Robert Doney, Physics Reports \textbf{462}, 21 (2008).
\bibitem{pgk_review} P.G. Kevrekidis, 
IMA Journal of Applied Mathematics, 389-423  (2011).
\bibitem{Theocharis_rev} G. Theocharis, N. Boechler, and C. Daraio,
in {\it Phononic Crystals and Metamaterials}, Ch. 6, 
Springer Verlag,
(New York, 2013).

\bibitem{Herbold09} E.B., Herbold, J. Kim,  V.F. Nesterenko, S. Wang, C. Daraio
Acta Mechanica, \textbf{205},  85-103 (2009).




\bibitem{Khatri} D. Khatri, D. Ngo, and C. Daraio, Granular Matter \textbf{14}, 63 (2012).
\bibitem{Li} F. Li, D. Ngo, J. Yang
and C. Daraio, Appl. Phys. Lett. \textbf{101}, 171903 (2012).
\bibitem{brillouin} L. Brillouin, Ann. Phys. {\bf 17}, 88 (1922).
\bibitem{diode1} B. Liang, X.S. Guo, J. Tu, D. Zhang, and J.C. Cheng, 
Nature Mater. {\bf 9}, 989 (2010).
\bibitem{diode2} N. Boechler, G. Theocharis and C. Daraio,
Nature Mater. {\bf 10}, 665 (2011).

\bibitem{Johnson} K. L. Johnson, Cambridge University Press (Cambridge, 1985).
\bibitem{Mathieu} Gertrude Blanch, Handbook of Mathematical Functions with Formulas, Graphs, and Mathematical Tables (Dover, New York, 1972).
\bibitem{Wei} Wei-Chau Xie, Dynamic Stability of Structures Cambridge University Press (Cambridge, 2006).
\bibitem{hill} W. Magnus and S. Winkler,
{\it Hill's equation}, Dover Publications (New York, 2004).

\bibitem{goldman} I. I. Gold'man and V. D. Krivchenkov, {\it Problems in
Quantum Mechanics} Dover Publications (New York, 1993).

\bibitem{Hussein} M. I. Hussein,  Proc. R. Soc. A  \textbf{465},  2825-2848 (2009).  

\bibitem{atkinson}
K. Atkinson, {\it An Introduction to
Numerical Analysis}, Wiley and Sons (New York, 1989).

\bibitem{Teschl}
G. Teschl, {\it Jacobi Operators and Completely Integrable Nonlinear Lattices},  Amer. Math. Soc.
(Providence, 2000).

\bibitem{gligoric} G. Gligori{\'c}, A. Maluckov, L. Hadzievski
and B.A. Malomed,
Chaos {\bf 24}, 023124 (2014).

\bibitem{pearl_pra}  P.J.Y. Louis, E.A. Ostrovskaya,
Yu.S. Kivshar, Phys. Rev. A {\bf 71}, 023612 (2005).

\bibitem{lax} P.D. Lax, {\it Linear Algebra}, J. Wiley \& Sons
(Hoboken, 2007).


\bibitem{dar01}
N. Boechler, G. Theocharis, S. Job, P. G. Kevrekidis, Mason A. Porter, and C. Daraio
Phys. Rev. Lett. {\bf 104}, 244302 (2010);
G. Theocharis, N. Boechler, P. G. Kevrekidis, S. Job, Mason A. Porter, and C. Daraio
Phys. Rev. E {\bf 82}, 056604 (2010);
C. Chong, P. G. Kevrekidis, G. Theocharis, and Chiara Daraio
Phys. Rev. E {\bf 87}, 042202 (2013);
C. Chong, F. Li, J. Yang, M. O. Williams, I. G. Kevrekidis, P. G. Kevrekidis, and C. Daraio
Phys. Rev. E {\bf 89}, 032924 (2014).

\end{thebibliography}
\end{document}